\documentclass[epj]{svjour}
\usepackage{graphicx} 
\usepackage{amsmath}
\usepackage{dcolumn}  
\usepackage{colordvi}
\usepackage{color}
\usepackage{epstopdf}
\usepackage{amssymb}
\usepackage{siunitx}
\usepackage{url}
\graphicspath{{ps}}
\usepackage[colorlinks=true,linkcolor=black,anchorcolor=black,citecolor=black,filecolor=black,menucolor=black,runcolor=black,urlcolor=black]{hyperref}
\usepackage{tabularx}
\usepackage{xspace}
\usepackage{authblk}
 
\usepackage{latexsym}
\usepackage{macros}

\begin{document}
\title{Hadronic Currents and Form Factors in three-body Semileptonic \Ptau Decays}
\author{Fabian Krinner\inst{1} \and Stephan Paul\inst{1,2}
}                     
\offprints{}          
\institute{Max Planck Institut f\"ur Physik, 80805 M\"unchen \and Physik-Department, Technische Universit\"at M\"unchen, 85748 Garching}
\date{Received: date / Revised version: date}
%
\abstract{
   Three-body semileptonic \Ptau-decays offer a path to understand the properties of
   light hadronic systems and CP symmetry violations through searches for electric dipole 
   moments. In studies of electro-weak physics, the hadronic part of the final 
   states has traditionally been described using the language of form factors. 
   Spectroscopic information, resolved in terms of orbital angular momentum quantum-numbers,
   is best being derived from an explicit decomposition of the hadronic current in the 
   orbital angular momentum basis. Motivated by the upcoming large data samples 
   from \PB factories, we present the full description of the hadronic currents decomposed 
   into quantum numbers of the hadronic final state using the isobar 
   picture. We present formulas for orbital angular momenta up to three and apply 
   the rules derived from hadron spectroscopy to formulate the decay chain of hadronic 
   three-body systems of arbitrary mass. We also translate this formalism to the language of 
   form factors and thereby correct insufficiencies found in previous 
   analyses of three-body hadronic final states.
} 
\maketitle

\section{Introduction}
\label{sec:introduction}
The upcoming large data samples from Belle and Belle 2 for \Ptau decays
will be an order of magnitude larger than previously analyzed samples, 
allowing for analyses of unprecedented detail.
The weak decay of the \Ptau lepton results in leptonic and semi-leptonic 
final states. The latter can serve as a clean source for hadronic systems 
produced from the vacuum. The weak current can either be of vector or axial 
vector nature. Restricting ourselves to light quarks only, we can view the 
\PW Boson to carry strong isospin $I=1$, as it couples to $u\bar d$ or 
$d\bar u$ \cite{burchat}. Based on arguments with conserved $G$-parity, 
odd numbers of pions in the final state can only be produced from the weak axial 
current, even ones from the weak vector current. Contributions through the weak 
vector current to odd numbers of pions are called second class 
currents~\cite{secondClass}.

The coupling of the weak axial-vector current to final states with odd 
numbers of pions can lead to quantum numbers  $J^P=1^+$ (axial-vector states) or 
$J^P=0^-$ (pseudo-scalar states) of the hadronic system. Indeed, the decay 
into a single pseudo-scalar ground-state pion is one of the dominating \Ptau final states 
(10.8\% branching fraction \cite{pdg}).  $J^P=1^+$ axial vectors on the other hand require 
at least 3 pions in the final state to constitute a first class current
not suppressed by $G$ parity\footnote{\JP denotes the spin $J$ 
of a state and \parity its eigenvalue under the parity transformation.}.

In this article, we focus on the semileptonic decay of the \Ptau lepton 
into three final state 
pions. Owing to the presence of both vector and axial-vector currents, \Ptau 
decays can a priori produce three-pion states with various combinations of 
quantum numbers \JP.

Although semileptonic \Ptau decays have been studied for the last 30 years, 
the structure of the hadronic current is still not understood sufficiently. 
Neither has the spectral distribution of the pseudo-scalar current been studied, 
nor have second class currents been found. In addition, the results on the 
decay branching for the \PaOne shows large deviations from results obtained 
by direct production in hadronic beam experiments, namely the contribution of 
the $(\Ppi\Ppi)_\Sw$-wave in the isobar seems much enhanced in Ref.~\cite{cleo} 
as compared to other experiments like in Refs.~\cite{Salvini:2004gz,Chung:2002pu}. 
The precise knowledge of the hadronic current, however, is a key requirement 
for determining the \Ptau magnetic moment through measurements of the \Ptau 
polarization or for searches for an electric dipole moment through spin 
correlations within the \Ptaupm-pair system.

All past analyses of the decay $\Ptaum\to3\Ppi\Pnu_\Ptau$ were restricted 
to the strong axial-vector component, which is dominantly passing through 
the \PaOne resonance. Possible additional contributions from pseudo-scalar 
resonances like the $\Ppi^\excited(1300)$ or from spin-exotic vector 
resonances like the $\PpiOne(1600)$, were assumed to have vanishing 
contributions and were neglected.

Hadronic systems within \Ptau decays are modeled by a hadronic current 
consisting of various contributions, called {\it partial waves}, that 
represent various hadronic resonances generating different orbital 
angular momenta in their decay chains. Omitting vector and pseudo-scalar 
partial waves in modelling the \Ptau decay to three pions, as done previously, 
led to a very limited set of seven partial waves. Partial waves 
describing decay chains including orbital angular momenta larger 
than two units of $\hbar$ were also neglected \cite{kuehn,cleo}.

The presence of partial waves describing $3\Ppi$ states other than the 
\PaOne not only allows to study their resonance nature, but also presents 
the opportunity to study the \PaOne itself in the absence of further 
hadronic interactions. Such interactions are present in  
alternative production mechanisms, like decays of heavy mesons or the 
production through strongly interacting hadrons. Possible additional 
resonances with different $J^P$ can serve as phase reference for the 
\PaOne, which improves constraints to the fit of the line shape as compared 
to a mere fit to the intensity distribution alone \cite{cleo,argus}.

The detailed study of axial-vectors, vectors, and pseudo-scalars in \Ptau-decays 
requires the derivation of previously omitted contributions to the hadronic 
current. We will construct these contributions for partial waves 
with appearing angular momenta of up to three units of $\hbar$ in their 
decay chain. In Sec.~\ref{sec:current}, we introduce the general structure 
of the amplitude and in Sec.~\ref{sec:tensor} we lay out the basic components 
to formulate hadronic currents. We give the explicit expressions in 
Secs.~\ref{sec:axial}, \ref{sec:vector}, and \ref{sec:scalar}. In Sec.~\ref{sec:formFactors}, we 
relate our findings to the more common language of form factors and compare 
to previous work in Sec.~\ref{sec:comparison}. In Sec.~\ref{sec:conclusion}, we will 
summarize and discuss the impact of the uncertainty on the hadronic current 
for the search of a tau lepton electric dipole moment.

\section{The hadronic current}
\label{sec:current}
The amplitude of the semileptonic decay of a \Ptau lepton into a 
neutrino and any hadronic final state is governed by the weak interaction 
and takes the form:
\begin{equation}
\mathcal{M} \propto \bar u_{\Pnu} \gamma_\mu \lr{1 - \gamma^5} u_{\Ptau} J^\mu_\had
,\end{equation}
where the left part describes the leptonic current, given by the corresponding 
Dirac spinors $u_{\Ptau}$ and $u_{\Pnu}$ and the Dirac matrices $\gamma$.
The hadronic current $J^\mu_\had$ is given by the final-state kinematics. 
The energy released in the decay can be shared among the neutrino 
and the hadronic final state resulting in a continuous spectrum of the invariant 
mass of the latter.
Since the hadronic current is governed by the 
strong interaction, there is no ab-initio calculation, but it is usually modeled 
as a sum of various contributions, which we will call {\it partial waves} from hereon:
\begin{equation}\label{eq:pwSum}
J^\mu_\had = \sum_{\pw\in\text{waves}} \mathcal{C}_\pw j_\pw^\mu
.\end{equation}
The sum extends  over a set of partial waves \pw describing the kinematics 
of the final-state particles. The complex-valued coefficients $\mathcal{C}_\pw$ 
encode 
the relative strengths and phases of the individual partial waves and 
are not known a priori but must be extracted from data via a partial-wave 
analysis. Even though this approach is applicable to all multi-body
hadronic final-states, here we will focus on three-particle final-states, 
particularly on $\Ptaum\to\Pnut + \Ppip\Ppim\Ppim$.

A partial wave corresponds to a particular description of the decay process,
starting with the production of a three-pion intermediate state $\PX$ 
with given $\JP_\PX$ quantum numbers. It also includes its decay into a two-pion 
intermediate state---the isobar \Pxi with given $\JP_\Pxi$ quantum numbers---and 
a single \Ppim with an orbital angular momentum \Lw with respect to the isobar 
and the final decay of the isobar into two pions $\Pxi\to\Ppip+\Ppim$. 
The quantum numbers $\JP_\PX$ and $\JP_\Pxi$ of the intermediate states 
\PX and \Pxi and the orbital angular momentum \Lw fully determine a 
partial wave \pw, which we will denote as
\begin{equation}\label{eq:naming}
\pw = \PX\LR{\Pxi\Ppi}_\Lw
.\end{equation}
In our case, the $\PaBase\LR{\PrhoBase\Ppi}_\Sw$ wave will 
dominate the spectral intensity distribution, which corresponds to a 
synthesis of the $\PaOne$ resonance decaying into a \Prho and a \Ppi in a 
relative \Sw wave and the subsequent decay of the \Prho into two 
charged pions in a relative \Pw wave.

The appearance of two undistinguishable $\Ppim$ 
in the final state $\Ppip_1\Ppim_2\Ppim_3$, 
requires the partial wave \pw of the hadronic current 
to be constructed observing Bose symmetry and thus contains 
two coherently summed parts:
\begin{equation}\label{eq:symm}
j_\pw^\mu = B_\PX\lr{s_{\PX}} \LR{B_\Pxi\lr{s_{(12)}} t^\mu_{(12),\pw} + B_\Pxi\lr{s_{(13)}} t^\mu_{(13),\pw} }
.\end{equation}
The real-valued tensor structures $t^\mu_{(ij),\pw}$ describe the 
two-particle combination $(ij)$ to form the isobar.
The complex-valued dynamic amplitudes $B_\PX\lr{s_\PX}$ and 
$B_\Pxi\lr{s_{(ij)}}$ describe the spectral distributions for 
$\PX$ and $\Pxi$ with respect to their invariant mass-squares 
$s_\PX$ and $s_{(ij)}$.

In contrast to the tensor structures $t^\mu_{(ij),\pw}$, the 
dynamic amplitudes cannot be derived from first principles, 
but have to be parameterized using existing knowledge.
The particular choice of a dynamic amplitude parameterization 
can be difficult and e.g. Refs.~\cite{cleo,isgur} 
use an elaborate parameterization including a mass function for the $\PaOne$.
However, as we will focus on the construction of the tensor structures 
$t^\mu_{(ij),\pw}$, we only give the most common parameterization, 
a Breit-Wigner amplitude with a given mass $m_\Pa$ and width 
$\Gamma_\Pa$, which in principle only describes an isolated 
resonance within a given partial wave laying far from any 
threshold\footnote{partial-wave analyses 
performed in kinematic bins of $s_\PX$ alleviate the 
necessity of a parameterization of $B_\PX\lr{s_{\PX}}$ 
and allow to extract this shape from data instead.}:
\begin{equation}\label{eq:boseDef}
B_\Pa\lr{s} = \frac{m_\Pa \Gamma_\Pa}{m_\Pa^2 - s - \complI m_\Pa \Gamma\lr{s}}
,\end{equation}
with
\begin{equation}
\Gamma\lr{s} = \Gamma_\Pa \lr{\frac{q\lr{s}}{q\lr{m_\Pa^2}}}^{2\ell +1} \frac{m_\Pa}{\sqrt{s}}
.\end{equation}
Here, $\ell$ is the orbital angular momentum appearing 
in the two body decay of the described resonance 
in its rest frame: $\PX\to\Pxi+\Ppim$ or $\Pxi\to\Ppip\Ppim$ 
(hereafter, the generic two-body decay will be dubbed as $\Pa\to\Pb+\Pc$).
The corresponding breakup momentum $q\lr{s = m_\Pa^2}$ in the decays of \PX and \Pxi 
has the generic form:
\begin{equation}\label{eq:breakup}
q\lr{s}=\sqrt{\frac{s^2 + m_{\Pb}^4 + m_{\Pc}^4 -2 \lr{s m_\Pb^2 + s m_\Pc^2 + m_\Pb^2 m_\Pc^2}}{4s}}
.\end{equation}

\section{Construction of hadronic tensors}
\label{sec:tensor}
We now construct the hadronic tensors 
$t^\mu_{(ij),\pw}$ defined in eq.~(\ref{eq:symm}) corresponding to the angular 
momentum quantum numbers in a given partial wave. From group theoretical 
considerations of the rotational group \SOthree
one can show, that an object with orbital angular momentum quantum 
number \Lw is described by a tensor, which is:
\begin{enumerate}
\item symmetric\label{it:sym}
\item traceless \label{it:tr}
\item of rank \Lw,\label{it:rank}
\item and transversal to the four-momentum of the decaying particle.\label{it:trans}
\end{enumerate}
In the rest frame of the decaying particle, the transversality condition 
translates to the tensor only having space-like components.
Rotations in the rest frame of a particle are governed by the 
rotational group \SOthree, subgroup of the full Lorentz group,
mixing only space-like components of tensors and leaving 
time components untouched. Thus, a tensor describing an object of 
non-zero spin cannot have time-like components in the corresponding 
rest frame. The requirements \ref{it:sym} and \ref{it:tr} are only 
relevant for an orbital angular momentum greater than one, since 
symmetry and trace are quantities defined only for tensors with rank 
greater than one.

We will begin with the construction of tensor amplitudes for the general 
two-body decay $\Pa\to\Pb+\Pc$. Energy and momentum conservation requires for their 
four-momenta to fulfill:
\begin{equation}\label{eq:kinabc}
p_\Pa^\mu = p_\Pb^\mu + p_\Pc^\mu
.\end{equation}
Based on these four-momenta we can now define the following 
objects\footnote{in the \Pa rest frame, $q_\Pa^\mu = \lr{q^0_\Pa,\vec q_\Pa}^\intercal$ 
with $\left|\vec q_\Pa\right| = q\lr{m_\Pa^2}$ as defined in eq.~(\ref{eq:breakup}).}:
\begin{equation}\label{eq:tensDef}
q_\Pa^\mu = \frac{1}{2}\lr{p_\Pb^\mu - p_\Pc^\mu};\ g_\Pa^{\mu\nu} = \metric^{\mu\nu} - \frac{p_\Pa^\mu p_\Pa^\nu}{s_\Pa};\ k_\Pa^\mu = g_\Pa^{\mu\nu}q_{\Pa,\nu}
,\end{equation}
where $s_\Pa=m_\Pa^2=p^\mu_\Pa p_{\Pa\mu}$ and $\metric^{\mu\nu}$ is the usual Minkowski 
metric. $g_\Pa^{\mu\nu}$ projects out the components of a four vector, which are 
transversal to $p_\Pa^\mu$. Thus, $k_\Pa^\mu$ is the only four-vector transversal 
to $p_\Pa^\mu$, which we can construct.

Using the objects defined in eq.~(\ref{eq:tensDef}), we now construct the tensor 
structures $\Tens^{\mu_1\ldots\mu_\Lw}_\Pa$, which describe the decay of 
the particle \Pa involving an orbital angular momentum \Lw, thus requiring \Lw Lorentz indices 
$\mu_1$ to $\mu_\Lw$.
The tensor for an \Sw-wave decay of orbital angular momentum zero $\Tens_\Pa$ is 
isotropic and thus simply given by unity.
The first non-trivial tensor represents a \Pw-wave decay and is 
given by the components $k_\Pa^\mu$ transversal to $p_\Pa^\mu$:
\begin{equation}\label{eq:spinOne}
\Tens_\Pa^\mu = k_\Pa^\mu
,\end{equation}
since $k_\Pa^\mu$ is the only vector we can construct 
from the available four-vectors in  eq.~(\ref{eq:kinabc}),
fulfilling all requirements listed above: $k^\mu_\Pa p_{\Pa,\mu} = 0$.

The next higher order tensor of interest is $\Tens_\Pa^{\mu\nu}$ of rank two 
corresponding to a \Dw-wave decay. It is given by
\begin{equation}\label{eq:spinTwo}
\Tens_\Pa^{\mu\nu} = \frac{3}{2} k_\Pa^\mu k_\Pa^\nu - \frac{1}{2} g_\Pa^{\mu\nu} \lr{k_\Pa^\varrho k_{\Pa,\varrho}}
,\end{equation}
which can be shown to fulfill all requirements listed above:
\begin{equation}
\Tens_\Pa^{\mu\nu} p_{\Pa\mu} = 0;\quad \Tens_\Pa^{\mu\nu}=\Tens_\Pa^{\nu\mu};\quad \Tens_{\Pa,\mu}^{\mu} = 0
.\end{equation}
An \Fw-wave decay is then described by the following rank 3 tensor:
\begin{align}
\Tens_\Pa^{\mu\nu\varrho} &= \frac52 k_\Pa^\mu k_\Pa^\nu k_\Pa^\varrho \\
& - \frac12 \lr{k_\Pa^\sigma k_{\Pa,\sigma}} \lr{k_\Pa^\mu g_\Pa^{\nu\varrho} + k_\Pa^\nu g_\Pa^{\mu\varrho} +k_\Pa^\varrho g_\Pa^{\mu\nu}} \nonumber
,\end{align}
which again can be shown to fulfil all requirements.
A recursive formula for the construction of higher orbital angular momenta \Lw
tensors $\Tens^{\mu_1\ldots\mu_\Lw}_\Pa$, with a number of \Lw Lorentz indices 
$\mu_1$ to $\mu_\Lw$ can be found in eq.${}^{\text{\cite{sarantsev}}}$~(9) of Ref.~\cite{sarantsev}.

\section{Axial vector currents}
\label{sec:axial}
Since the weak interaction violates parity, it proceeds via 
both vector and axial-vector currents, with $\JP = 1^-$ 
and $\JP = 1^+$. Since the axial-vector current is not 
strictly conserved, the corresponding component in the 
weak interaction may give rise to contributions to the 
hadronic current with quantum numbers $\JP_\PX = 0^-$ 
and $\JP_\PX = 1^+$, the latter of which is expected to 
dominate in the three-pion channel. Thus, we begin by 
constructing its contributions to the hadronic current. 

Using the tensors defined in Sec.~\ref{sec:tensor}, we
can construct the non-symmetrized hadronic currents 
$t^\mu_{(ij),\pw}$ for a partial wave \pw as defined 
in eq.~(\ref{eq:symm}) and with particles $(ij)$ forming 
the isobar. The most simple axial wave is 
$\PaBase\LR{\PfZero\Ppi}_\Pw$, with the \PaOne quantum 
numbers being $\JP_{\PaBase} = 1^+$. Here, the only 
non-zero angular momenta are the spin of the \PaOne
and the orbital angular momentum $\Lw = 1$ enclosed 
by \PfZero and \Ppim\footnote{\PfZero denotes a generic 
isobar resonance with $\JP_\Pxi = 0^{+}$.}. Thus, we find:
\begin{equation}\label{eq:fZeroP}
t^\mu_{(12),\PaBase\LR{\PfZero\Ppi}_\Pw} = \Tens^\mu_{\PX}
,\end{equation}
defined in eq.~(\ref{eq:spinOne}), with $\PX= \PaOne$. As 
the \PfZero decays into two particles labeled 1 and 2, 
we denote the tensor in eq.~(\ref{eq:fZeroP}) with (12) such that 
$p_\Pb^\mu = p_{(12)}^\mu$ and $p_\Pc^\mu = p_3^\mu$, as 
defined in Sec.~\ref{sec:tensor}. 

The next partial wave we discuss is $\PaBase\LR{\PrhoBase\Ppi}_\Sw$, 
which describes the decay of an \PaOne axial vector meson into the 
\Prho and \Ppim enclosing a relative \Sw wave. 
Here, the only non-zero angular momentum is given by the isobar spin, since 
the final state pions $\Ppip_1$ and $\Ppim_2$ are spinless and thus their relative orbital angular 
momentum must equal the isobar spin. 
The corresponding tensor is $\Tens^\mu_{(12)}$, defined in eq.~(\ref{eq:spinOne}) with $p_\Pb^\mu = p_1^\mu$ 
and $p_\Pc^\mu = p_2^\mu$. This tensor is transversal to $p_{(12)}^\mu$ by 
definition, but not transversal to $p_{\PX}^\mu$. To ensure this, we need to 
apply $g_{\PX}^{\mu\nu}$, given in eq.~(\ref{eq:tensDef}), and obtain:
\begin{equation}\label{eq:rhoPiS}
t^\mu_{(12),\PaBase\LR{\PrhoBase\Ppi}_\Sw} = g_{\PX}^{\mu\nu} \Tens_{(12),\nu}
.\end{equation}

Since the spin of the \PaOne and the \Prho can also 
couple to $\Lw=2$, we now also consider the partial wave $\PaBase\LR{\PrhoBase\Ppi}_\Dw$.
The orbital angular momentum of $\Lw =2$ is represented by $\Tens_{\PX}^{\mu\nu}$ given in 
eq.~(\ref{eq:spinTwo}), with $p_\Pb^\mu = p_{(12)}^\mu$              
and $p_\Pc^\mu = p_3^\mu$. The decay of \Prho is described by the same 
structure as in eq.~(\ref{eq:rhoPiS}) and we find for the total decay chain:
\begin{equation}
t^\mu_{(12),\PaBase\LR{\PrhoBase\Ppi}_\Dw} = \Tens^{\mu\nu}_{\PX} \Tens_{(12),\nu}
,\end{equation}
which is transversal to $p_{\PX}^\mu$ by definition.

Finally we consider the wave $\PaBase\LR{\PfTwo\Ppi}_\Pw$.
The appearance of an orbital angular momentum $\Lw= 1$ is
 described by $\Tens^\mu_{\PX}$, like in eq.~(\ref{eq:fZeroP}) 
and the decay of the \PfTwoT isobar is described by 
$\Tens^{\mu\nu}_{(12)}$. As discussed for the decay of the 
\Prho, the orbital angular momentum between $\Ppip_1$ and 
$\Ppim_2$  must match the isobar spin. Again, 
$\Tens^{\mu\nu}_{(12)}$---defined in eq.~(\ref{eq:spinTwo}) 
with $p_\Pb^\mu = p_1^\mu$ and $p_\Pc^\mu = p_2^\mu$---is 
not transversal to $p_{\PX}^\mu$, so we use 
$g_{\PX}^{\mu\nu}$ to obtain:
\begin{equation}
t^\mu_{(12),\PaBase\LR{\PfTwo\Ppi}_\Pw} = g_{\PX}^{\mu\nu} \Tens_{(12),\nu\varrho}\Tens^\varrho_{(12)}
.\end{equation}

For a general amplitude with arbitrary \Lw we need an isobar spin of $J_\Pxi = \Lw\pm1$ to 
be able to construct a spin 1 tensor. For $\Lw = J_\Pxi +1$ we find
\begin{equation}
t^\mu_{(12),\PaBase\LR{\Pxi_j\Ppi}_\Lw} = \Tens_{\PX}^{\mu\nu_1\ldots\nu_{J_\Pxi}} \Tens_{(12),\nu_1\ldots\nu_{J_\Pxi}}
,\end{equation}
since the only non-contracted Lorentz index $\mu$ comes from $\Tens_{\PX}^{\mu\nu_1\ldots\nu_{J_\Pxi}}$,
which is already transversal to $p_\PX^\mu$. This is not the case for $\Lw = J_\Pxi -1$, where we 
need to apply the projection operator and find:
\begin{equation}
t^\mu_{(12),\PaBase\LR{\Pxi_j\Ppi}_\Lw} = g_{\PX}^{\mu\nu} \Tens_{(12),\nu\varrho_1\ldots\varrho_\Lw} \Tens_{\PX}^{\varrho_1\ldots\varrho_\Lw}
.\end{equation}

For the Bose symmetrized tensors $t^\mu_{(13),a}$, defined in eq.~(\ref{eq:symm}), 
the four-momenta $p_2^\mu$ and $p_3^\mu$ are interchanged in all formulas. Even though
we only explicitly mentioned the ground-state \PaOne resonance and the \Prho and \PfTwoT 
isobars, the formulas in this section are equally valid for all excited $\PaBase^\excited$, 
$\PrhoBase^\excited$, $\PfZero^\excited$ and $\PfTwo^\excited$ resonances.

\section{Vector currents}
\label{sec:vector}
Since the vector current is conserved, the vector component of the weak 
interaction can only give rise to a hadronic vector current with quantum 
numbers $\JP_\PX = 1^-$ and not $\JP_\PX = 0^+$. Owing to $G$-parity, the 
hadronic vector current is only expected to contribute to final states 
with an even number of final-state pions, while its contributions to the 
three-pion final state is usually assumed to vanish. In this section, we 
nevertheless give the expected dominant contribution to a hadronic three-pion 
vector current to be able to explicitly search for isospin violation.

The most prominent example of this is the $\PpiOne\LR{\PrhoBase\Ppi}_\Pw$ wave, 
which involves the spin exotic state \PpiOneM with $\JP_{\PpiOne} = 1^-$. In 
this wave, the isobar spin and the orbital angular momentum are equal $J_\Pxi = \Lw = 1$
 and thus we cannot obtain a tensor of spin 1 by 
simple contraction of Lorentz indices as in Sec.~\ref{sec:axial}. Using the totally 
antisymmetric tensor $\antiEps^{\mu\nu\varrho\sigma}$, we construct:
\begin{equation}\label{eq:exoticTensor}
t^\mu_{(12),\PpiOne\LR{\PrhoBase\Ppi}_\Pw} = \antiEps^{\mu\nu\varrho\sigma} p_{\PX,\nu} \Tens_{\PX,\varrho} \Tens_{(12),\sigma}
,\end{equation} 
which is transversal to $p_{\PX}^\mu$ by definition, due to the permutation 
properties of $\antiEps^{\mu\nu\varrho\sigma}$. The tensors of orbital 
angular momentum 1, $\Tens_{\PX}^\mu$
and $\Tens_{(12)}^\mu$, are given in eqs.~(\ref{eq:fZeroP}) and (\ref{eq:rhoPiS}).
Due to the antisymmetric tensor $\antiEps^{\mu\nu\varrho\sigma}$, 
the tensor $t^\mu_{(12),\PpiOne\LR{\PrhoBase\Ppi}_\Pw}$ transforms 
differently under the parity operator than the tensors constructed 
in Sec.~\ref{sec:axial} and therefore the corresponding three-pion state 
is of type \PpiOne instead of type \PaBase.

For the construction of vector currents involving higher isobar spins 
$J_\Pxi$ and angular momenta $\Lw$, both must be equal and all but one of the
Lorentz indices on both corresponding tensors must be contracted with one 
another. The single remaining index of both tensors is contracted 
with $\antiEps^{\mu\nu\varrho\sigma}$, similar to eq.~(\ref{eq:exoticTensor}).

\section{Current for pseudo-scalars}
\label{sec:scalar}
As mentioned in Sec.~\ref{sec:axial}, the non-conservation of the 
axial-vector current in the weak interaction also allows
pseudo-scalar resonances with $\JP_\PX=0^-$ to contribute to the 
hadronic current.
In analogy to the decay of \Ptau leptons into a single pion, 
which also carries $\JP_\Ppi=0^-$, 
the pseudo-scalar hadronic current is simply given by $p_\PX^\mu$ multiplied by a 
Lorentz scalar $t_{(12),\Ppi^\excited\LR{\Pxi\Ppi}_\Lw}$:
\begin{equation}\label{eq:scalarGeneral}
t^\mu_{(12),\Ppi^\excited\LR{\Pxi\Ppi}_\Lw} = p_{\PX}^\mu\, t_{(12),\Ppi^\excited\LR{\Pxi\Ppi}_\Lw}
,\end{equation}
where the spin $J_\Pxi$ of the isobar and \Lw must be equal. The scalar
factor $t_{(12),\Ppi^\excited\LR{\Pxi\Ppi}_\Lw}$ for the decay 
of a pseudo-scalar \PX via an isobar \Pxi can be constructed from 
the same tensors, as the axial and vector currents:
\begin{equation}
t_{(12),\Ppi^\excited\LR{\Pxi\Ppi}_\Lw} = \Tens_{\PX}^{\mu_1\ldots\mu_\Lw} \Tens_{(12),\mu_1\ldots\mu_\Lw}
.\end{equation}
Note, that in contrast to axial and vector currents, this 
pseudo scalar current does not have to be transversal to $p_{\PX}^\mu$, 
since it must be invariant under rotations in the \PX rest frame 
and therefore can have only a time component.

\section{Connection to form factors}
\label{sec:formFactors}
We may also use a formulation for the hadronic current, which 
is complementary to eq.~(\ref{eq:pwSum}), namely an expansion in 
terms of the four four-vectors constructed from the four-momenta 
of the final state particles. This expansion introduces 
complex-valued form factors to describe the hadronic structure. 
These four-vectors are:
\begin{align}
&p_{i\perp}^\mu = g_{\PX}^{\mu\nu}p_{i,\nu}\nonumber\\
&p_{\antiEps}^\mu = \antiEps^{\mu\nu\lambda\sigma}p_{1,\nu} p_{2,\lambda} p_{3,\sigma}\label{eq:ffVectors}\\
&p_{\PX}^\mu = p_1^\mu + p_2^\mu + p_3^\mu\nonumber
,\end{align}
where $p_i^\mu$ are the pion-momenta and $g_{\PX}^{\mu\nu}$ is 
defined in eq.~(\ref{eq:tensDef}). Thus, $p_{i\perp}^\mu$ are the 
components of $p_i^\mu$ that are transversal to $p_\PX^\mu$.
Using these, the expansion of the had\-ro\-nic current in terms of 
form factors $F_\ff$ reads:
\begin{equation}\label{eq:currentDec}
J^\mu_\had = p_{2\perp}^\mu F_2 + p_{3\perp}^\mu F_3 + p_{\antiEps}^\mu F_\antiEps + p_{\PX}^\mu F_\text{s}
,\end{equation}
where $F_2$ and $F_3$ are two axial form-factors, $F_{\antiEps}$
is the vector form-factor, and $F_\text{s}$ is the scalar from 
factor. The form factors are functions of the Lorentz invariant 
quantities $\{s_\PX, s_{(12)}, s_{(13)}, s_{(23)}, m_\Ppi^2\}$. 
Similar decompositions into form factors of the had\-ro\-nic current 
have already been given in Refs.~\cite{kuehn,cleo} to which we 
relate our results in Sec.~\ref{sec:comparison}.

Since there is only one four-vector for each of 
the scalar and vector form-factors with the respective 
transformation properties, only a single scalar form-factor
and a single vector form-factor appear.
There are two independent axial form-factors, since 
there are only two linearly independent corresponding four-vectors due 
to momentum conservation in the \PX rest frame:
\begin{equation}\label{eq:zeroSum}
p_{1\perp}^\mu =- p_{2\perp}^\mu - p_{3\perp}^\mu
.\end{equation}

Analogue to the hadronic current, we can expand the form factors
into a series of partial waves:
\begin{align}\label{eq:parWavForFac}
F_\ff = \sum_{\pw \in \text{waves}}\mathcal{C}_\pw B_\PX\lr{s_{\PX}}\Big[&B_\Pxi\lr{s_{(12)}}f_{\ff,\pw}^{(12)} \\
	+& B_\Pxi\lr{s_{(13)}}f_{\ff,\pw}^{(13)}\Big] \nonumber
.\end{align}
The sum extends over the sub-set of partial waves $\pw$ with 
$\JP_\PX$ quantum numbers matching the corresponding $F_\ff$. 
Partial-wave form-factors $f_{\ff,\pw}^{(ij)}$ encode the contribution of 
partial wave \pw to the form factor $\ff\in\{2,3,\antiEps,\text{s}\}$.
$(ij)$ denote the final-state particles forming the isobar. 
In the following, we will match the tensor expressions derived in 
Secs.~\ref{sec:axial}--\ref{sec:scalar} to the corresponding 
form factors in eq.~(\ref{eq:parWavForFac}).

\subsection{Axial form factors}
\label{sec::axialFF}
Here, we discuss the expansion of the axial currents from Sec.~\ref{sec:axial}
for a partial wave \pw in terms of partial-wave form-factors $f^{(12)}_{\ff,\pw}$, 
such that:
\begin{equation}\label{eq:pwff}
t_{(12),\pw}^\mu = p_{2\perp}^\mu f^{(12)}_{2,\pw} + p_{3\perp}^\mu f^{(12)}_{3,\pw} 
.\end{equation}
For the axial-vector like waves discussed in Sec.~\ref{sec:axial}, 
we derive the corresponding form factors by performing all Lorentz-contractions 
given in Sec.~\ref{sec:axial}:
\begin{equation}\label{eq:ffZeroP}
f_{2,\PaBase\LR{\PfZero\Ppi}_\Pw}^{(12)} = 0; \quad f_{3,\PaBase\LR{\PfZero\Ppi}_\Pw}^{(12)} = -1
\end{equation}
and
\begin{equation}\label{eq:fRhoPiS}
f_{2,\PaBase\LR{\PrhoBase\Ppi}_\Sw}^{(12)} = -1; \quad f_{3,\PaBase\LR{\PrhoBase\Ppi}_\Sw}^{(12)} = -\frac12
.\end{equation}
For the $\PaBase\LR{\PrhoBase\Ppi}_\Dw$ wave, the respective form factors are:
\begin{align}
f_{2,\PaBase\LR{\PrhoBase\Ppi}_\Dw}^{(12)} =& \frac{1}{8}\lr{2 s_{(12)} - s_{\PX} + 2m_\Ppi^2}-\frac{\lr{s_{(12)} - m_\Ppi^2}^2}{8s_{\PX}};\label{eq:fRhoPiD}\\
f_{3,\PaBase\LR{\PrhoBase\Ppi}_\Dw}^{(12)} =& \frac{3}{16}\lr{s_{(13)} - s_{(23)}}\lr{1 + \frac{s_{(12)} - m_\Ppi^2 }{s_{\PX}}} \nonumber \\
+&  f_{2,\PaBase\LR{\PrhoBase\Ppi}_\Dw}^{(12)}\big/2 \nonumber
\,.\end{align}
And for $\PaBase\LR{\PfTwo\Ppi}_\Pw$:
\begin{align}
f_{2,\PaBase\LR{\PfTwo\Ppi}_\Pw}^{(12)} =& \frac{3}{16}\lr{s_{(13)} - s_{(23)}}\lr{ 1 + \frac{s_{(12)} - m_\Ppi^2}{s_{\PX}}};\label{eq:fFFPiP}\\
f_{3,\PaBase\LR{\PfTwo\Ppi}_\Pw}^{(12)} =&\frac{4m_\Ppi^2 - s_{(12)}}{32s_{\PX}s_{(12)}} \lr{s_{\PX} + s_{(12)} - m_\Ppi^2}^2 \nonumber\\
+& f_{2,\PaBase\lr{\PfTwo\Ppi}_\Pw}^{(12)}/2 \nonumber
\,.\end{align}
Note, that eqs.~(\ref{eq:fRhoPiS})--(\ref{eq:fFFPiP}) are only 
valid in the case of identical final state masses 
\footnote{we used $p_i^\mu p_j^\mu = \frac{s_{(ij)} - 2 m_\Ppi^2}{2}$ in the 
calculation.}
$m_i^2 = p_i^\mu p_{i,\mu} = m_\Ppi^2$, while the results 
of Sec.~\ref{sec:axial} are valid in full generality. 
For arbitrary final-state masses, we give the form factors
in Sec.~\ref{sec:masses}. The Bose symmetrized axial 
form-factors are obtained as:
\begin{equation}
f_{2,\pw}^{(13)} = \left.f_{3,\pw}^{(12)}\right|_{s_{(12)}\leftrightarrow s_{(13)}}\text{and}\quad f_{3,\pw}^{(13)} = \left.f_{2,\pw}^{(12)}\right|_{s_{(12)}\leftrightarrow s_{(13)}}
.\end{equation}
In Sec.~\ref{sec:masses}, we also give form factors for the $\PaBase\LR{\PfTwo\Ppi}_\Fw$ and the
$\PaBase\LR{\PrhoBase_3\Ppi}_\Dw$ waves, the latter containing the spin~3 resonance $\PrhoThreeM$.
\subsection{Vector form-factor}
For the vector current, we can write the tensor amplitude derived in 
Sec.~\ref{sec:vector} as:
\begin{equation}
t^\mu_{(ij), \PpiOne\LR{\PrhoBase\Ppi}_\Pw} = p_\varepsilon^\mu f_{\varepsilon,\PpiOne\LR{\PrhoBase\Ppi}_\Pw}^{(ij)}
,\end{equation}
for which we find
\begin{equation}
f_{\antiEps,\PpiOne\LR{\PrhoBase\Ppi}_\Pw}^{(12)} = 1;\quad f_{\antiEps,\PpiOne\LR{\PrhoBase\Ppi}_\Pw}^{(13)} = -1
.\end{equation}
\subsection{Scalar form-factors}
Since the scalar part of the hadronic current already contains the 
desired structure for the decomposition into form-factors we 
simply identify $f_{\text{s},\pw}^{(12)}=t_{(12),\pw}$ defined in 
eq.~(\ref{eq:scalarGeneral}). Here, we give explicit expressions 
for $\Ppi^\excited\LR{\PfZero\Ppi}_\Sw$ and 
$\Ppi^\excited\LR{\PrhoBase\Ppi}_\Pw$:
\begin{equation}
f_{\text{s}, \Ppi^\excited\LR{\PfZero\Ppi}_\Sw}^{(12)} = 1
\end{equation}
and 
\begin{equation}\label{eq:scalarRhoFF}
f_{\text{s}, \Ppi^\excited\LR{\PrhoBase\Ppi}_\Pw}^{(12)} = \frac18 \lr{s_{(23)} - s_{(13)}}\lr{1 + \frac{s_{(12)} - m_\Ppi^2}{s_{\PX}}}
.\end{equation}
Note, that eq.~(\ref{eq:scalarRhoFF}) again only holds for equal final-state masses.
The formulas for arbitrary final state masses and for the $\Ppi^\excited\LR{\PfTwo\Ppi}_\Dw$-
and $\Ppi^\excited\LR{\PrhoBase_3\Ppi}_\Fw$-waves
are given in Sec.~\ref{sec:masses}. The Bose symmetrized scalar form-factors are:
\begin{equation}
f_{\text{s},\pw}^{(13)} = \left.f_{\text{s},\pw}^{(12)}\right|_{s_{(12)}\leftrightarrow s_{(13)}}
\end{equation}

\section{Comparison with previous work}
\label{sec:comparison}
In this section, we relate our findings to  previously 
published work in Refs.~\cite{kuehn} and \cite{cleo}.

In Ref.~\cite{kuehn}, the decomposition of the hadronic 
currents is given in eq.${}^{\text{\cite{kuehn}}}$~(3), 
using four form factors 
labeled $F_i^\text{\cite{kuehn}}$ modifying the vectors
$V_i^{\text{\cite{kuehn}}\mu}$ built from a combination 
of the four momenta of the outgoing three-particle state\footnote{
since our work uses similar notation to previous work, we annotate 
quantities defined outside this article with the respective literature 
reference to avoid confusion.}.
The vector $V_4^{\text{\cite{kuehn}}\mu}$ corresponds to 
our $p_\PX^\mu$ and we can identify $F_4^\text{\cite{kuehn}}$ 
with our scalar form-factor $F_\text{s}.$

The ordering of final-state particles in Ref.~\cite{kuehn}
is given by $\Ppim_1\Ppim_2\Ppip_3$ and thus we identify the 
corresponding momenta $q_i^{\text{\cite{kuehn}}\mu}$ with our momenta $p_i^\mu$ such,
that:
\begin{equation}
q_1^{\text{\cite{kuehn}}\mu} = p_2^\mu;\quad q_2^{\text{\cite{kuehn}}\mu} = p_3^\mu;\quad q_3^{\text{\cite{kuehn}}\mu} = p_1^\mu
.\end{equation}
This corresponds to two permutations in the definition 
of $V_3^{\text{\cite{kuehn}}\mu}$ as compared to
eq.~(\ref{eq:ffVectors}). We can 
identify the vector form-factor as $F_\antiEps = iF_3^\text{\cite{kuehn}}$.
The remaining vectors in eq.${}^{\text{\cite{kuehn}}}$~(3) of Ref.~\cite{kuehn} are:
\begin{equation}
V_1^{\text{\cite{kuehn}}\mu} = p_{2\perp}^\mu - p_{1\perp}^\mu;\quad V_2^{\text{\cite{kuehn}}\mu} = p_{3\perp}^\mu - p_{1\perp}^\mu
.\end{equation}
Using eq.~(\ref{eq:zeroSum}), we can relate our
axial form-factors $F_2$ and $F_3$ to the axial 
form-factors $F_1^\text{\cite{kuehn}}$ and $F_2^\text{\cite{kuehn}}$
of Ref.~\cite{kuehn}, as follows:
\begin{equation}
F_2 = 2 F_1^\text{\cite{kuehn}} + F_2^\text{\cite{kuehn}};\quad F_3 = F_1^\text{\cite{kuehn}} + 2F_2^\text{\cite{kuehn}}
.\end{equation}

In Ref.~\cite{cleo}, the 
decomposition of the hadronic current is given in 
eq.${}^{\text{\cite{cleo}}}$~(A2) of Ref.~\cite{cleo}. 
The authors of Ref.~\cite{cleo} do not eliminate one of the 
axial form-factors using eq.~(\ref{eq:zeroSum}) and thus use a 
total of five form factors.
Ref.~\cite{cleo} discusses the decay into $\Ppiz_1\Ppiz_2\Ppim_3$,
thus we translate it to the $3\Ppipm$ channel using the 
particle ordering $\Ppim_1\Ppim_2\Ppip_3$, consistent with 
Ref.~\cite{kuehn}. Using this ordering, we identify:
\begin{equation}
F_\text{s} = c_4 F_4^\text{\cite{cleo}};\quad F_\antiEps = ic_5 F_5^\text{\cite{cleo}}
.\end{equation}
The axial part $J^\mu_\text{axial}$  of the hadronic current of Ref.\cite{cleo} can be written as:
\begin{align} J^\mu_\text{axial} &=
c_1\lr{p_{3\perp}^\mu - p_{1\perp}^\mu} F_1^\text{\cite{cleo}}  \\\nonumber
+&c_2\lr{p_{1\perp}^\mu - p_{2\perp}^\mu} F_2^\text{\cite{cleo}} + 
c_3\lr{p_{2\perp}^\mu - p_{3\perp}^\mu} F_3^\text{\cite{cleo}}
.\end{align}
And we identify the form factors $F_2$ and $F_3$ from eq.~(\ref{eq:zeroSum}):
\begin{align}
F_2 &= c_1 F_1^\text{\cite{cleo}} -2 c_2F_2^\text{\cite{cleo}} + c_3 F_3^\text{\cite{cleo}}\\
F_3 &= 2c_1F_1^\text{\cite{cleo}} - c_2F_2^\text{\cite{cleo}} - c_3F_3^\text{\cite{cleo}}
.\end{align}
The Lorentz invariant quantities contained in the form 
factors match as:
\begin{equation}
s_1 = s_{(13)}; \quad s_2 = s_{(12)};\quad s_3 = s_{(23)}
.\end{equation}

We also compare the results for the individual partial wave 
contributions to the hadronic current given in Sec.~\ref{sec:axial} 
above with the expressions
given in eq.~${}^{\text{\cite{cleo}}}$(A3). First, we note
that Ref.~\cite{cleo} does not include a parameterization 
for the dynamic amplitude of the $3\Ppi$ state \PX, 
since the analysis was performed independently 
in mass bins of \PX, $m_{3\Ppi} = \sqrt{s_\PX}$. Second, in the dynamic 
amplitudes in Ref.~\cite{cleo} an additional factor $F_{Rj}\lr{k_k}$
multiplies the isobar Breit-Wigner\footnote{The nominal fit of Ref.~\cite{cleo} has 
$F_{Rj}\lr{k_k} = 1$, making it equal to our approach.}.
Since this factor only depends 
on the invariant masses of \PX and \Pxi, 
it can be absorbed into the dynamic amplitudes introduced in 
eq.~(\ref{eq:boseDef}). As stated in Sec.~\ref{sec:current}, we do not 
discuss this factor.

Finally, we compare the tensor structures for the individual 
partial waves. 
The mathematical formulation of the amplitudes corresponds to the 
$\Ppiz\Ppiz\Ppim$ final state and thus Bose symmetrization is already 
explicitly written in Ref.~\cite{cleo} for $\PrhoBase$-like
isobars, since these can be formed by both possible $(\Ppim\Ppiz)$
systems. However, no symmetrization is needed for $\PfZero$-like and $\PfTwo$-like 
isobars, since these can only be formed by $(\Ppiz\Ppiz)$. 
For our final state $\Ppip\Ppim\Ppim$, all isobars appear with a 
Bose symmetrization term. 

We now compare the Bose-symmetrization terms for 
the individual partial waves.
For the $\PaBase\LR{\PfZero\Ppi}_\Pw$ 
wave, which corresponds to waves 6 and 7 in Ref.~\cite{cleo} our 
results agree with eq.${}^{\text{\cite{cleo}}}$~(A3). 

For the $\PaBase\LR{\PrhoBase\Ppi}_\Sw$ waves, which correspond to 
waves 1 and 2 in Ref.~\cite{cleo}, we only agree in 
case of equal-mass mesons forming the isobar. 
Thus, for $\Ppiz\Ppiz\Ppim$ a small deviation is introduced, 
owing to the difference between $m_{\Ppipm}$ and $m_{\Ppiz}$.
The origin of this deviation is a missing projection operator 
$g_{(12)}^{\mu\nu}$, as defined in eq.~(\ref{eq:tensDef}), coming 
from the \Prho propagator \cite{feindt}.

For $\PaBase\LR{\PrhoBase\Ppi}_\Dw$ waves, there is a discrepancy 
with respect to the corresponding waves 3 and 4 of Ref.~\cite{cleo}. 
This might not be surprising at first sight, as their amplitudes 
do not correspond to \Lw eigenstates, but to two Born term amplitudes. 
Using the naming scheme of eq.~(\ref{eq:naming}), replacing the upper case letters for 
\Lw by lower case letters and thus following Ref.~\cite{cleo}, we 
can write these Born term amplitudes as given in eq.${}^\text{\cite{feindt}}$~(1)
of Ref.~\cite{feindt}:
\begin{align}
t^\mu_{(12),\PaBase\LR{\PrhoBase\Ppi}_s} &= g_\PX^{\mu\nu}\ \eta_{\nu\varrho}\ k_{(12)}^\varrho, \\
t^\mu_{(12),\PaBase\LR{\PrhoBase\Ppi}_d} &= g_\PX^{\mu\nu}\ p_{(12),\nu}\ p_{\PX\varrho}\ k_{(12)}^\varrho 
.\end{align}
Ref.~\cite{feindt} states, that every linear combination of these 
two Born term amplitudes with Lorentz scalar coefficients constitutes 
a valid amplitude for the decay $\Ptau\to3\Ppipm+\Pnu$. Indeed, we can 
write the amplitudes corresponding to \Lw eigenstates as such linear 
combinations of Born term amplitudes. For our \Dw-wave amplitude we find\footnote{
The \Lw eigenstates in Ref.~\cite{feindt} correspond to the ones in the 
helicity formalism e.g. layed out in Ref.~\cite{helicity}, while we use 
the covariant tensor formalism described in Ref.~\cite{sarantsev}. The connection 
between both formalisms can be found in Ref.~\cite{friedrich}.
}:
\begin{align}
t^\mu_{(12),\PaBase\LR{\PrhoBase\Ppi}_\Dw} =& c_s\lr{s_\PX,s_{(12)}} t^\mu_{(12),\PaBase\LR{\PrhoBase\Ppi}_s}\\\nonumber
 +& c_d\lr{s_\PX,s_{(12)}} t^\mu_{(12),\PaBase\LR{\PrhoBase\Ppi}_d}
,\end{align}
where $c_s\lr{s_\PX,s_{(12)}}$ and $c_d\lr{s_\PX,s_{(12)}}$ are two Lorentz-scalar 
coefficients that, however, depend on $s_\PX$ and $s_{(12)}$. We can invert this to 
obtain the Born term amplitudes as linear combination of the amplitudes 
describing \Lw eigenstates in an analogue way:
\begin{align}\label{eq:linComb}
t^\mu_{(12),\PaBase\LR{\PrhoBase\Ppi}_d} =-&\frac{c_s\lr{s_\PX,s_{(12)}}}{c_d\lr{s_\PX,s_{(12)}}} t^\mu_{(12),\PaBase\LR{\PrhoBase\Ppi}_\Sw} \\\nonumber
+& \frac{1}{c_d\lr{s_\PX,s_{(12)}}}t^\mu_{(12),\PaBase\LR{\PrhoBase\Ppi}_\Dw}
.\end{align}
We now insert the Born term $\PaBase\LR{\PrhoBase\Ppi}_d$ tensor structure 
into eq.~(\ref{eq:symm}) to construct the corresponding partial-wave hadronic-current 
within the isobar model and find using eq.~(\ref{eq:linComb}):
\begin{equation}\label{eq:bornCurrent}
\begin{array}{r@{}l}
j_{\PaBase\LR{\PrhoBase\Ppi}_d}^\mu =& B_{\PaBase}\lr{s_\PX}B_{\PrhoBase}\lr{s_{(12)}} \Big(-\frac{c_s\lr{s_\PX,s_{(12)}}}{c_d\lr{s_\PX,s_{(12)}}} t^\mu_{(12),\PaBase\LR{\PrhoBase\Ppi}_\Sw}\\
                                    &+ \frac{1}{c_d\lr{s_\PX,s_{(12)}}}t^\mu_{(12),\PaBase\LR{\PrhoBase\Ppi}_\Dw}\Big) + \text{Bose symm.}
\end{array}
\end{equation}
The coefficients in eq.~(\ref{eq:linComb}), which depend on $s_\PX$ and 
$s_{(12)}$, now result in additional terms multiplying the partial-wave 
hadronic-currents in eq.~(\ref{eq:bornCurrent}), expressed in the 
\Lw-eigenstate basis. Thus, the naive use of Born term amplitudes 
within the context of the isobar model results in effective distortions 
of the dynamic isobar amplitudes $B_{\PaBase}\lr{s_\PX}$ and 
$B_{\PrhoBase}\lr{s_{(12)}}$.

We also find discrepancies for the $\PaBase\LR{\PfTwo\Ppi}_\Pw$ wave, 
again resulting in an $s_\PX$ and $s_{(12)}$ dependent factor, which can
be absorbed into the dynamic amplitudes, leading to distortions in the 
corresponding two-hadron dynamic amplitude. We attribute the 
discrepancy to a missing projector for the \Pw-wave tensor, which is 
due to the amplitudes of Ref.~\cite{cleo} not corresponding to 
\Lw-eigenstates. And as Ref.~\cite{cleo} does not include the 
$\PaBase\LR{\PfTwo\Ppi}_f$, amplitudes for \Lw eigenstates cannot 
be constructed as linear combinations.

Finally, additional discrepancies between our approach and
Ref.~\cite{cleo} appear for the case of final-states formed by 
non-equal mass particles for the $\PaBase\LR{\PrhoBase\Ppi}_\Dw$-wave 
and the $\PaBase\LR{\PfTwo\Ppi}_\Pw$-wave. This is again due 
to a missing projection operator for the isobar, as in the 
$\PaBase\LR{\PrhoBase\Ppi}_\Sw$-wave.

\section{Summary and discussion}
\label{sec:conclusion}
In this article, we introduced the general formalism
to describe the hadronic current in semileptonic 
\Ptau decays within the framework of the isobar model. 
We explicitly constructed individual partial wave 
contributions using the decay $\Ptaum\to3\Ppipm+\Pnu$ 
as an example.

For this formalism, we included contributions from 
partial waves with orbital angular momenta up to three 
units of $\hbar$, which had not been accounted for in 
previous works. In addition to contributions from 
axial-vector currents, we also allowed for contributions 
from vector and pseudo-scalar resonances, which were 
assumed to vanish in Refs.~\cite{kuehn,cleo}. These 
additional contributions allow for the search of 
pseudo-scalar $3\Ppi$ resonances, like the $\Ppi(1300)$, 
and of vector resonances, like the $\Ppi_1(1600)$ in 
\Ptau decays.

We then translated the individual contributions of the 
partial-wave hadronic currents into the formalism of 
form factors and compared to previous results from 
Refs.~\cite{kuehn,cleo}.

Finally, we confront our results for the individual
partial wave currents to the ones quoted in 
Ref.~\cite{cleo} and we find two sources of 
discrepancies. First,  it is important to note that 
the waves in Ref.~\cite{cleo} are not describing 
\Lw-eigenstates and thus do not trivially translate 
to the isobar model, as they cause distortions of 
the dynamic amplitudes. Second, the waves in 
Ref.~\cite{cleo} do not project out the transversal 
components of the tensor structure describing the isobar 
decay.This leads to discrepancies with respect to our 
approach for the case of final states with non-equal 
mass particles.

We want to state, that an explicit construction of 
partial waves with given quantum numbers is necessary 
to make results interpretable in terms of the hadronic 
excitation spectrum. A fit with partial waves with mixed 
quantum numbers and distorted dynamic amplitudes does not 
give reliable information on the physics sorted by orbital 
angular momentum quantum numbers in \Ptau decays. 

The expressions developed here allow for the first time 
to extend partial-wave analyses of the decay 
$\Ptau\to3\Ppim+\Pnut$, by including previously neglected 
contributions: higher orbital angular momenta with $L>2$, 
as well as pseudo-scalar and vector contributions.

Beyond the study of the hadronic spectrum itself, a proper 
description of the decay kinematics of the decay 
$\Ptau\to3\Ppipm+\Pnu$ is necessary for the precision 
measurements of \Ptau properties like the electric dipole 
moment (EDM). For this, we have to correlate the decay 
distributions of both \Ptaupm leptons in \Ptau-pair 
production. For the measurement of an EDM we single out 
CP odd correlations and search for CP violations therein.
We thereby rely on maximal sensitivity to the spin 
correlation of the \Ptaupm pair. The necessary spin 
analyzing power is best obtained from hadronic final states 
\cite{nachtmann}. If we use three-pion final states we need 
to know the contributions of the different partial waves for 
we have to precisely describe the $\Ptau\to3\Ppi+\Pnu$ decay 
distribution. Using three pions in the final state, we rely 
on the proper description of the $\Ptau\to3\Ppipm + \Pnu$ 
decay kinematics reflecting the contributions of the partial 
waves.


We have investigated the sensitivity of the value 
for the EDM extracted from pseudo data on the 
description used for the $3\Ppi$ final state. For this, 
we have first generated pseudo data using various 
input values for the EDM and then analyzed the 
data using the method from Ref.~\cite{nachtmann}. 
For the analysis, we assumed a variety of partial 
wave compositions and determined the deviation of 
the extracted EDM values from the input values.
In particular, the use of a wrong hadronic model 
leads to an under-estimation of the EDM by a 
factor of up to two. We found these deviations to 
depend on the overlap between the input hadronic 
model and the model used in the analysis. We thus 
demonstrated that a false EDM may be extracted from 
data, which is generated through an incomplete 
understanding of the hadronic final states. 

\appendix
\section{Form factors for arbitrary final-state masses}
\label{sec:masses}
To complete the discussion in Sec.~\ref{sec:formFactors}, we extend the 
formulae for the form factors to three-hadron final states with arbitrary 
masses, $m_1$, $m_2$ and $m_3$. The isobar shall be formed by particles 
1 and 2, as denoted in (12). The form factors for the wave $\PaBase\LR{\PfZero\Ppi}_\Pw$ 
are described in eq.~(\ref{eq:ffZeroP}). To obtain the proper expression for 
$\PaBase\LR{\PrhoBase\Ppi}_\Sw$ we express all Lorentz contractions in the 
tensor formalism (see Sec.~\ref{sec:tensor}) by Lorentz invariants:
\begin{equation}
f_{2,\PaBase\LR{\PrhoBase\Ppi}_\Sw}^{(12)} = -1;\quad f_{3,\PaBase\LR{\PrhoBase\Ppi}_\Sw}^{(12)} = -\frac{1}{2}\lr{1 - \frac{m_1^2 - m_2^2}{s_{(12)}} }
.\end{equation}
In order to extend the formulation of form factors for partial waves 
involving higher oribital angular momenta,
we define the following expressions for convenience:
\begin{eqnarray}
\text{FI}   = &-s_\PX^2 -\lr{m_3^2 - s_{(12)}}^2 + 2 s_\PX\lr{s_{(12)} + m_3^2}\\
\text{FII}  = &\lr{s_\PX + s_{(12)} -m_3^2}\Big( m_2^2 m_3^2 + s_{(12)} \lr{m_2^2+m_3^2} \nonumber\\
 &   - s_{(12)}^2 + s_\PX \lr{s_{(12)} - m_2^2} \\
 &+ m_1^2\lr{s_\PX - m_3^2 + s_{(12)}} - 2 s_{(12)} s_{(13)}\Big) \nonumber\\
\text{FIII} = & s_\PX s_{(12)} \lr{m_1^2 - m_2^2}^2 \\
 &- s_\PX s_{(12)}^2\lr{2 m_1^2 + 2m_2^2 - s_{(12)}}\nonumber \\
\text{FIV}  = & m_1^4 + \lr{s_{(12)} - m_2^2}^2 - 2m_1^2\lr{s_{(12)} + m_2^2}\\
\text{FV}   =& \lr{s_\PX -m_3^2} \lr{m_1^2 - m_2^2} \\
&- s_{(12)} \lr{s_{(13)} - s_{(23)}} \nonumber
.\end{eqnarray}
For the $\PaBase\LR{\PrhoBase\Ppi}_\Dw$ wave, $f_{2,\PaBase\LR{\PrhoBase\Ppi}_\Dw}^{(12)}$ 
stays unchanged and is given in  eq.~(\ref{eq:fRhoPiD}), where $m_\Ppi^2$ is substituted with $m_3^2$. 
For $f_{3,\PaBase\LR{\PrhoBase\Ppi}_\Dw}^{(12)}$ we find:
\begin{eqnarray}
f_{3,\PaBase\LR{\PrhoBase\Ppi}_\Dw}^{(12)} = &\frac{1}{16 s_\PX s_{(12)}}\Big[\lr{s_{(12)} - m_1^2 + m_2^2}\text{FI} \\
&- 3\lr{s_\PX - m_3^2 + s_{(12)}}\text{FV}\Big]\nonumber
\end{eqnarray}
For the generalized version of the $\PaBase\LR{\PfTwo\Ppi}_\Pw$ form factors, we find:
\begin{equation}
f_{2,\PaBase\LR{\PfTwo\Ppi}_\Pw}^{(12)} = -3\frac{s_\PX +s_{(12)}- m_3^2}{16s_\PX s_{(12)}}\text{FV}
\end{equation}
and
\begin{equation}
\begin{array}{r@{}l}
f_{3,\PaBase\LR{\PfTwo\Ppi}_\Pw}^{(12)} =&\frac{s_{(12)} - m_1^2 + m_2^2}{2 s_{(12)}}f_{2,\PaBase\LR{\PfTwo\Ppi}_\Pw}^{(12)} + \frac{\text{FIII}}{32s_\PX^2 s_{(12)}^3} \\
 &\cdot \Big[s_\PX \lr{3s_{(12)} - s_\PX + m_3^2} - 4s_\PX s_{(12)}\\
& + \lr{m_3^2 - s_{(12)}}\lr{s_\PX - m_3^2 + s_{(12)}}\Big]
\end{array}
\end{equation}
For completeness, we also 
give the from factors for orbital angular momenta of three. For the $\PaBase\LR{\PfTwo\Ppi}_\Fw$ wave, 
we find:
\begin{equation}
f_{2,\PaBase\LR{\PfTwo\Ppi}_\Fw}^{(12)} =  \frac{3\text{FI}\cdot\text{FII}}{64s_\PX^2 s_{(12)}}
\end{equation}
and
\begin{equation}
\begin{array}{r@{}l}
f_{3,\PaBase\LR{\PfTwo\Ppi}_\Fw}^{(12)} =& \frac{1}{256 s_\PX^2 s_{(12)}^2}\Big[- 3 \text{FI}\cdot\Big( 4 \text{FIII}+\text{FV}^2\\& - 2\lr{s_{(12)} - m_1^2 + m_2^2} \text{FII} \Big)
\\&+2\text{FIV}\cdot\text{FI}^2 -15 \text{FII}^2 \Big].
\end{array}
\end{equation}
The form factors for the $\PaBase\LR{\PrhoBase_3\Ppi}_\Dw$ are given by:
\begin{equation}
\begin{array}{r@{}l}
f_{2,\PaBase\LR{\PrhoBase_3\Ppi}_\Dw}^{(12)} = &\frac{1}{256 s_\PX^2 s_{(12)}^2} \Big[-3\lr{s_\PX - m_3^2 + s_{(12)}}^2 \text{FIV}\cdot\text{FI}\\ 
&-15 \text{FII}^2 + \text{FI} \cdot\Big(-20\text{FIII} - 5\text{FV}^2\Big)+ \text{FIV}\\
&\cdot\lr{16s_\PX s_{(12)} + \lr{s_\PX - m_3^2 + s_{(12)}}^2} \Big]
\end{array}
\end{equation}
and
\begin{equation}
\begin{array}{r@{}l}
f_{3,\PaBase\LR{\PrhoBase_3\Ppi}_\Dw}^{(12)} =& -\frac{1}{512 s_\PX^3 s_{(12)}^4}\Big[\Big( 10 \text{FV}^2 \lr{s_{(12)} - m_1^2 + m_2^2}\\
&\cdot s_\PX s_{(12)}  + 4\text{FIII} \text{FV} \cdot\lr{s_\PX - m_3^2 + s_{(12)}} \Big)\\
& \lr{ \lr{s_\PX  + s_{(12)} - m_3^2}^2 + 2 s_\PX s_{(12)}}\\ 
+3 \text{FI}\cdot &\text{FIII} \lr{s_{(12)} - m_1^2 + m_2^2} \lr{s_\PX - m_3^2 + s_{(12)}}^2 \\
+ \text{FIII}\cdot&\lr{s_{(12)} - m_1^2 + m_2^2}\\
& \lr{\lr{s_\PX + m_3^2 - s_{(12)}}^2 - 4 s_\PX m_3^2}^2\Big]
.\end{array}
\end{equation}
Below we give the scalar form factors for arbitrary final-state 
masses. For the $\Ppi^\excited \LR{\PfZero\Ppi}_\Sw$ wave, we again find $f^{(12)}_{\text{s},\Ppi^\excited \LR{\PfZero\Ppi}_\Sw}$ ${= 1}$ and
for the $\Ppi^\excited\LR{\PrhoBase\Ppi}_\Pw$ we find:
\begin{equation}
f^{(12)}_{\text{s},\Ppi^\excited \LR{\PrhoBase\Ppi}_\Pw} = -2/3f_{2,\PaBase\LR{\PfTwo\Ppi}_\Pw}^{(12)}
.\end{equation}
The scalar form factors for partial waves involving \PfTwo and $\PrhoBase_3$ isobars
are given by:
\begin{equation}
\begin{array}{r@{}l}
f^{(12)}_{\text{s},\Ppi^\excited\LR{\PfTwo\Ppi}_\Dw} =& \frac{1}{256s_\PX^2 s_{(12)}^2}\Big[
  -\Big(s_\PX^2 - 2 s_\PX \lr{m_3^2 - 5 s_{(12)}}\\&
 + \lr{m_3^2 - s_{(12)}}^2 -3\lr{s_\PX - m_3^2 + s_{(12)}}^2\Big) \\&
\cdot\text{FIV} \cdot\text{FI}  +9 \lr{s_\PX - m_3^2 + s_{(12)}}^2 \text{FV}^2 \\&
+3 \text{FI}\Big( 4\text{FIII}  + \text{FV}^2\Big)\Big]
\end{array}
\end{equation}
and
\begin{equation}
\begin{array}{r@{}l}
f^{(12)}_{\text{s},\Ppi^\excited\LR{\PrhoBase_3\Ppi}_\Fw} =& \frac{\text{FII}}{200 s_\PX^3 s_{(12)}^3}
\Big[  15 \lr{s_\PX - m_3^2 + s_{(12)}}^2 \text{FIV}\cdot \text{FI} 
\\&      -3\text{FIV} \cdot  \text{FI} \lr{-3\text{FI} + 20 s_\PX s_{(12)}} 
\\& +  25 \text{FII}^2  + 15 \text{FI}     \lr{ 4 \text{FIII} +  \text{FV}^2}\Big]
.\end{array}
\end{equation}
Since in contrast to axial-vector and pseudo-scalar contributions, 
vector contributions are suppressed by $G$-parity, we do not give 
explicit expressions for vector form factors involving higher 
orbital angular momenta.

%

\end{document}